\documentclass[a4paper]{article}
\usepackage{INTERSPEECH2021}

\usepackage{cite}
\usepackage{url}
\usepackage{graphicx}
\usepackage{hyperref}
\usepackage{verbatim}
\usepackage{multirow}
\usepackage[table,xcdraw]{xcolor}
\usepackage{array}
\usepackage{tabularx}
\newcolumntype{Y}{>{\centering\arraybackslash}X}

\title{Unified Source-Filter GAN: Unified Source-filter Network Based On Factorization of Quasi-Periodic Parallel WaveGAN}
\name{Reo Yoneyama$^1$, Yi-Chiao Wu $^1$, Tomoki Toda$^2$}
\address{
  $^1$Nagoya University Furo-cho,Chikusa-ku,Nagoya,464–8601 Japan\\
  $^2$Information Technology Center,Nagoya University Furo-cho, Chikusa-ku,Nagoya, 464–8601 Japan}
\email{$^1$\{yoneyama.reo, yichiao.wu\}@g.sp.m.is.nagoya-u.ac.jp, $^2$tomoki@icts.nagoya-u.ac.jp}

\begin{document}

\maketitle
\begin{abstract}

We propose a unified approach to data-driven source-filter modeling using a single neural network for developing a neural vocoder capable of generating high-quality synthetic speech waveforms while retaining flexibility of the source-filter model to control their voice characteristics. Our proposed network called unified source-filter generative adversarial networks (uSFGAN) is developed by factorizing quasi-periodic parallel WaveGAN (QPPWG), one of the neural vocoders based on a single neural network, into a source excitation generation network and a vocal tract resonance filtering network by additionally implementing a regularization loss. Moreover, inspired by neural source filter (NSF), only a sinusoidal waveform is additionally used as the simplest clue to generate a periodic source excitation waveform while minimizing the effect of approximations in the source filter model. The experimental results demonstrate that uSFGAN outperforms conventional neural vocoders, such as QPPWG and NSF in both speech quality and pitch controllability.

\end{abstract}

\section{Introduction}

Currently, neural vocoders \cite{wavenet, pwn, pwg, samplernn, fftnet, flowavenet, melgan, multi-melgan, waveffjord, hinet, waveglow, hooligan, vocgan, pwngan, sa_pwngan, pap_gan, glotgan_2017, glotgan_2019, wavernn, clarinet, nsf_2019, nsf_2020, lpcnet, nhv, gelp, glotnet} usually achieve very high-fidelity speech generation by directly modeling raw speech waveforms using advanced neural networks without ad hoc designs. On the other hand, because of the data-driven nature and unified network architecture \cite{qpnet, qppwg}, the speech components controllability of the neural vocoders are usually inferior to the conventional source-filter vocoders \cite{straight, world}. Therefore, it is desired to develop a neural vocoder capable of high-fidelity and controllable speech generation.

To improve the controllability, there have been proposed many generation models integrating conventional parametric-based source-filter models with deep neural network architectures \cite{nsf_2019, nsf_2020, lpcnet, nhv, gelp, glotnet}. For example, neural source-filter (NSF) \cite{nsf_2019, nsf_2020} realizes speech generation based on non-autoregressive modeling by non-linear filtering of parametrically generated source excitation signals with multiple dilated convolutional layers. LPCNet \cite{lpcnet} adopts a WaveRNN \cite{wavernn}-like architecture to generate residual signals while a linear filtering process is applied to generate speech waveforms as in the conventional linear predictive coding (LPC) vocoder \cite{lpc_1982, lpc_1995}. Generative adversarial network (GAN) based neural homomorphic vocoder (NHV) \cite{nhv} first develops neural-based linear-time-variant (LTV) filters with the input pulse trains and white noise to generate mixed source excitations, and then a trainable causal finite impulse response (FIR) filter is applied to the excitations for generating the output waveforms. Although these hybrid neural vocoders have successfully improved the controllability by integrating parametric-based approaches, the synthetic speech quality of these vocoders tends to be inferior to that of data-driven unified neural vocoders. Moreover, there is still room for improvements in the controllability. 

To achieve high-fidelity and high-controllability speech generation, we propose a GAN-based framework to introduce the source-filter model with fewer ad hoc designs into a single neural network. The generator is designed by factorizing quasi-periodic Parallel WaveGAN (QPPWG) \cite{qppwg} into two cascaded networks corresponding to the source excitation generation and resonance filtering, and these two networks are jointly optimized in the training stage. Only a sinusoidal waveform is additionally used as the simplest clue to generate a periodic source excitation waveform while minimizing the effect of approximations in the source filter model. Moreover, to generate reasonable source excitation signals, an additional auxiliary loss is applied to the source excitation network.
The main contributions of this paper are summarized as follows:
\begin{itemize}
  \item We propose a unified framework for neural vocoders attaining an interpretable and tractable source-filter-like architecture, making it possible to well models excitation generation and resonance filtering while keeping the simplicity of training.
  \item The proposed method achieves better fundamental frequency ($F_0$) controllability than the conventional neural vocoders, such as QPPWG and NSF, while attaining high-fidelity speech generation even in $F_0$ transformation scenarios.
\end{itemize}

\section{Related work}

This chapter describes non-AR neural vocoders: Parallel WaveGAN (PWG) \cite{pwg}, QPPWG \cite{qppwg}, and NSF \cite{nsf_2019}.
PWG and QPPWG are the basis of our method, and we use QPPWG as one of the baseline methods. NSF is a semi-parametric neural vocoder based on the source-filter model, which is used as the other baseline method.

\subsection{Parallel WaveGAN (PWG)}

PWG is a GAN-based method for generating raw waveforms. It is a compact model without an autoregressive structure or a causal mechanism, and can achieve fast speech generation with high fidelity.
The model consists of two networks, generator (G) and discriminator (D). The WaveNet-based generator, which is conditioned by auxiliary features, learns to make the discriminator recognize the generated sample as $real$. This process can be written as follows:
\begin{equation}
    \mathcal{L}_{adv}(G, D) = \mathbb{E}_{\bm{z} \sim \mathcal{N}(0,I)} \left[ (1-D(G(\bm{z})))^2 \right],
\end{equation}
where $\bm{z}$ is random noise distributed from Gaussian distribution. Note that all auxiliary features of the generator are omitted in this paper for simplicity.

The discriminator learns to identify the generated sample as $fake$ and the natural sample as $real$. This process can be written as follows:
\begin{equation}
    \begin{split}
    \mathcal{L}_{D}(G, D) = \mathbb{E}_{\bm{x} \sim p_{data}} \left[ (1-D(\bm{x}))^2 \right]&
    \\ + \mathbb{E}_{\bm{z} \sim \mathcal{N}(0,I)} \left[ D(G(\bm{z}))^2 \right]&,
    \end{split}
\end{equation}
where $\bm{x}$ denotes the natural samples and $p_{data}$ denotes the
data distribution of the natural samples. PWG also adopts multi-STFT loss \cite{pwg} as an auxiliary loss $\mathcal{L}_{aux}(G)$ to improve the training stability. In conclusion, the final loss function of the generator can be written as a weighted sum of $\mathcal{L}_{aux}$ and $\mathcal{L}_{adv}$ as follows:
\begin{equation}
    \mathcal{L}_{G}(G, D) = \mathcal{L}_{aux}(G) + \lambda_{adv} \mathcal{L}_{adv}(G, D)
\end{equation}
where $\lambda_{adv}$ is hyperparameter for weight and empirically set to 4.0 in this paper.


\subsection{Quasi-periodic parallel WaveGAN (QPPWG)}

Although PWG achieves high-fidelity speech generation, the fully data-driven nature makes PWG lack the explicit controllability of each speech component especially when unseen auxiliary features are given such as $F_0$ outside the $F_0$ range of training data. To alleviate this issue, in QPPWG pitch-dependent dilated convolution neural networks (PDCNNs), which dynamically changes dilation sizes adapting to pitch, are introduced to PWG. PDCNN facilitates QPPWG to capture the very long-term dependencies of periodic components and makes the pitch of the QPPWG-generated speech more consistent with the auxiliary $F_0$.

For the ordinary dilated convolution neural networks, the dilation size $d$, is predefined and time-invariant. The dilation sizes $d_t$ of PDCNNs are dynamically defined at each time step $t$ as follows:
\begin{equation}
    d_t = d \times f_s / (f_{0,t} \times a) 
\end{equation}
where $f_s$ is the sampling rate, $f_{0,t}$ is an $F_0$ value at time $t$, and $a$ is a hyperparameter called dense factor, which determines the sparsity of the PDCNNs, and empirically set to 4.0 in this paper.

\subsection{Neural source-filter (NSF)}

NSF is a neural vocoder based on source-filter model, and divided into three modules: a condition module, a source module, and a filter module. An $F_0$ sequence $f_{0,1}, \cdots, f_{0,T}$ and a spectral feature sequence (e.g. mel-spectrogram) are used as an input of NSF. The condition module upsamples the input features and extracts feature embeddings for the resonance filtering. In the source module, by treating $f_{0,t}$ as the instantaneous frequency, the fixed number of sinusoidal wave basis signals is generated, where the $h$-th basis signal $e_t^{(h)}$ is given by
\begin{equation} \label{math:nsf-sine}
    e_t^{(h)} = 
    \begin{cases}
        \sin \left(\displaystyle{ \sum_{k=1}^{t} } 2 \pi \frac{h f_{0,k}}{f_s} + \phi \right) + n_t & \mbox{if} ~ f_{0,t} > 0 \\
        \displaystyle{\frac{1}{3 \sigma} n_t} & \mbox{if} ~ f_{0,t} = 0,
    \end{cases}
\end{equation}
where $f_s$ is sampling frequency, $\phi \in [-\pi, \pi]$ is a random initial phase, $n_t \sim \mathcal{N}(0,\sigma^2)$ is a Gaussian noise, and $h$ is the scale factor of the fundamental frequency. These signals are merged using a feed-forward network to output the source excitations. The filter module modulates the source signal using multiple stages of dilated convolution and affine transformations similar to those in ClariNet \cite{clarinet}. NSF adopts multi-resolution STFT loss to learn the difference between the output and the target waveform in the spectral domain. Unlike that of PWG, it calculates the mean square error (MSE) for the log power spectrum.

\section{Proposed method: unified source-filter GAN (uSFGAN)}

Our proposed method, uSFGAN is based on QPPWG, but differs on the generator in several ways: (1) the generator is explicitly split into a source-network and a filter-network; (2) a sinusoidal signal is used as an additional input of the source-network, and (3) a regularization term for the output of the source-network is added to the auxiliary loss. On the other hand, the discriminator is the same as that of QPPWG.

\subsection{Network architecture}

\begin{figure}
\begin{center}
    \includegraphics[width=7.8cm]{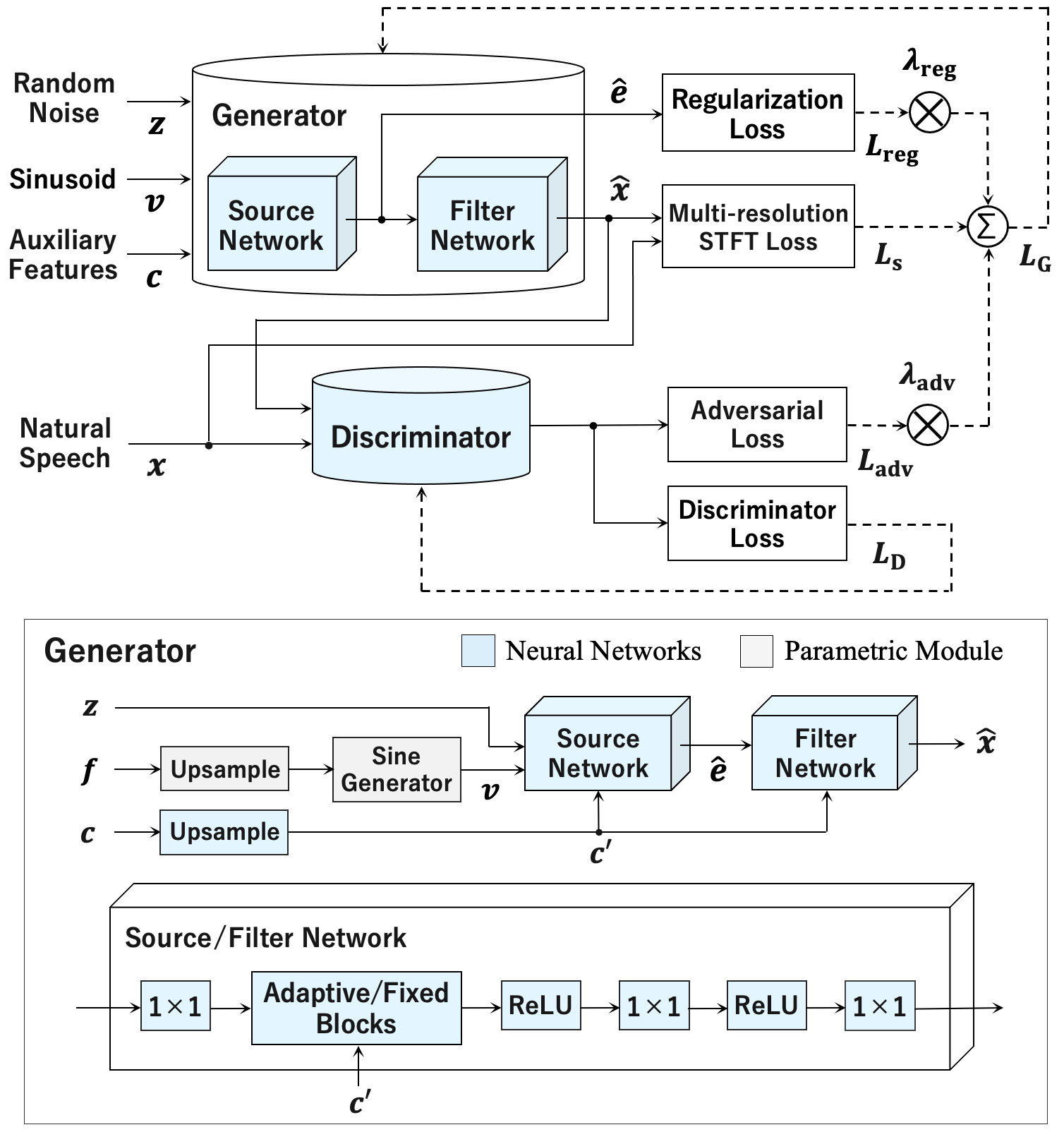}
    \vspace{-2mm}
    \caption{Architecture of the proposed method, uSFGAN.}
    \label{fig:architecture}
\end{center}
\vspace{-8mm}
\end{figure}

As the proposed architecture shown in Fig.\ref{fig:architecture}, the generator of uSFGAN receives random noise $\bm{z}$ sampled from Gaussian distribution, an $F_0$ sequence $\bm{f}$, and an auxiliary feature sequence $\bm{c}$ as the input, where $\bm{f}$ and $\bm{c}$ are assumed to be extracted per frame. A sinusoidal signal $\bm{v} = v_1, \cdots, v_T$ is first generated on the basis of upsampled $\bm{f}$ as follows:
\begin{equation} \label{math:sine}
    v_t = 
    \begin{cases}
        \sin \left(\displaystyle{ \sum_{k=1}^{t} } 2 \pi \frac{f_{0,k}}{f_s} \right)& if ~ f_{0,t} > 0 \\
        0 & if ~ f_{0,t} = 0,
    \end{cases}
\end{equation}
where $f_{0,t}$ is the instantaneous frequency at time $t$, and $f_s$ is the sampling frequency. Unlike QPPWG adopting only noise inputs, the sinusoidal signal input is used to make the estimation of the harmonic components easier and improve the learning efficiency of the proposed source-network. Then, $\bm{v}$ is combined with $\bm{z}$ as a two-channel input of the source-network. The source-network performs the pitch-dependent dilated convolution conditioned on the upsampled auxiliary features $\bm{c}$ to output the source excitation signal $\hat{\bm{e}}$. The generated source excitation signal is used as the input to the filter-network and is also used to calculate the spectral envelope regularization loss, which is used for the auxiliary loss, as mentioned in Section \ref{ssec:auxloss}. In the filter-network, non-causal dilated convolution with fixed dilation sizes is performed. The output waveform is input to the discriminator and also used to calculate the multi-resolution STFT auxiliary loss.

\subsection{Spectral envelope regularization loss}
\label{ssec:auxloss}

To encourage the source-network to output a reasonable source excitation signal, one constraint is imposed on the output of the source-network. As in the traditional source-filter vocoders, such as STRAIGHT \cite{straight} and WORLD \cite{world}, we assume that the spectral structure of the source excitation signal consists of harmonic components and stochastic components, and its spectral envelope is flat, {\em i.e.}, the power of spectral envelope is constant over all frequency. We adopt regularization to satisfy this assumption on the spectral envelope of the source excitation signal in the output of the source-network.

We use a simplified algorithm of cheaptrick \cite{cheaptrick} to extract the spectral envelope from the output signal of source-network. The original algorithm of cheaptrick is composed of three steps: (1) $F_0$ adaptive windowing and calculation of log power spectrum, (2) $F_0$ adaptive smoothing in the spectral domain, and (3) $F_0$ adaptive liftering in the cepstrum domain. In order to speed up the spectral envelope estimation process, we apply several modifications to the cheaptrick algorithm. First, we directly use the $F_0$ values given as the auxiliary feature $\bm{f}$ rather than extracting $F_0$ values from the output signal. Those $F_0$ values are further rounded to integers, and then, the corresponding windows and liftering functions are obtained in advance. Moreover, the step (2) is omitted because this process requires a relatively large processing time and the $F_0$ adaptive spectral envelope extraction is still performed by the $F_0$ adaptive liftering in the step (3). Although these modification causes slight degradation of the spectral envelope estimation accuracy, it doesn't cause any significant issues as the precise spectral envelope estimation is not necessary for the regularization.

The spectral envelope regularization loss is given by
\begin{equation} \label{math:reg aux loss}
    \mathcal{L}_{reg}(G) = \frac{1}{2}\sum_{n=1}^{N} \sum_{k=1}^{K} \hat{E}_k^{(n)2},
\end{equation}
where $\hat{E}_k^{(n)}$ is the $k$-th frequency component of log power spectral envelope extracted from $\bm{\hat{e}}$ at $n$-th time frame by the simplified cheaptrick algorithm. Note that when this loss reaches to 0, linear power values of the spectral envelope are 1 over all frequency and time frames.

\subsection{Training criteria}

The same adversarial losses as QPPWG but different auxiliary losses are adopted in uSFGAN training. Our method uses two types of auxiliary losses: multi-resolution STFT loss and the spectral envelope regularization loss. The STFT loss is defined as follows: 
\begin{equation} \label{math:multi-resolution stft loss}
    \mathcal{L}_s(G) = \frac{1}{2} \sum_{n=1}^{N} \sum_{k=1}^{K} \left[ \log \frac{\mbox{Re}(Y_k^{(n)})^2 + \mbox{Im}(Y_k^{(n)})^2}{\mbox{Re}(\hat{Y}_k^{(n)})^2 + \mbox{Im}(\hat{Y}_k^{(n)})^2} \right] ^ 2,
\end{equation}
where $Y_k^{n}$ and $\hat{Y}_k^{n}$ are the $k$-th STFT component at the $n$-th time frame of a natural waveform and the output waveform. $\mbox{Re}$ and $\mbox{Im}$ denote real part and imaginary part, respectively. This STFT loss is different from that of PWG \cite{pwg}, but the same as that of NSF \cite{nsf_2019, nsf_2020}. 
Finally, our auxiliary loss is represented as follows:
\begin{equation} \label{math:our aux loss}
    \mathcal{L}_{aux}(G) = \frac{1}{M}\sum_{m=1}^{M}\mathcal{L}_s^{(m)}(G) + \lambda_{reg} \mathcal{L}_{reg}(G),
\end{equation}
where $M$ is the number of STFT losses using various STFT parameters, and $\lambda_{reg}$ is a hyperparameter balancing the two auxiliary losses and empirically set to 1.0 in this paper. 

\section{Experimental evaluation}

\subsection{Experimental conditions}

To investigate the effectiveness of our proposed method, we compared four different models: publicly available pretrained NSF (hn-sinc-nsf-9 \cite{hn-sinc-nsf-9}) model referred to as PT-NSF, NSF with WORLD features referred to as WORLD-NSF, QPPWG, uSFGAN, and uSFGAN without the spectral envelope regularization loss. We adopted $F_0$ conversion to evaluate the controllability.

For the training data, we used 4000 utterances from CMU-ARCTIC database \cite{arctic} consisting of more than 1000 utterances each from four speakers: slt, bdl, clb, and rms. We used a set of 264 utterances consisting of 66 utterances from each speaker as validation data and another set of 264 utterances as test data. The sampling frequency was set to 16000 Hz by down-sampling.

WORLD-NSF used the same architecture as PT-NSF. QPPWG used 10 adaptive blocks and 10 fixed blocks as the optimized setting. uSFGAN used 30 adaptive blocks for the source-network, and 30 fixed blocks for the filter-network. uSFGAN was trained with the RAdam optimizer \cite{radam} ($\epsilon = 10^{-6}$) with 400 k iterations as in QPPWG. The generator of uSFGAN is trained with only auxiliary loss in the first 100 k iterations, and then trained with the adversarial loss as well as the auxiliary loss in the remaining 300 k iterations. The parameter settings of the multi-resolution STFT loss are shown in Table \ref{table:stft loss params}.

\begin{table}[tb]
\caption{Parameter settings for multi-resolution STFT loss. We apply a hanning window before the FFT process.}
\vspace{-2mm}
\label{table:stft loss params}
\fontsize{8.0pt}{9.6pt}
\selectfont
{%
\begin{tabularx}{\columnwidth}{Yrrr}
\toprule
\multicolumn{1}{c}{\textbf{STFT loss}} & \multicolumn{1}{c}{\textbf{Frame shift}} & \multicolumn{1}{c}{\textbf{Frame size}} & \multicolumn{1}{c}{\textbf{DFT bins}} \\ \midrule
$L_{s}^{(1)}$ & 80 (5 ms)   & 320 (20 ms)   & 512 \\
$L_{s}^{(2)}$ & 40 (2.5 ms) & 80 (5 ms)     & 128 \\
$L_{s}^{(3)}$ & 640 (40 ms) & 1920 (120 ms) & 2048 \\ \bottomrule
\end{tabularx}%
}
\vspace{-5mm}
\end{table}

In PT-NSF, $F_0$ extracted by YAAPT \cite{yaapt} and mel-spectrum extracted by STFT-based method were used for the input. In the other models, $F_0$, spectral envelope, and aperiodicity extracted by WORLD \cite{world} were used. The window length was set to 64 ms and the shift length was set to 5 ms in all models. The spectral envelope was parameterized into 25-dimensional mel-cepstral coefficients, and aperiodicity was coded into 1-dimension. The unvoiced/voiced intervals were represented as a binary feature. For PT-NSF, the target speech used for training was normalized after feature extraction. On the other hand, for the other models, no normalization process was applied to the target speech.

\subsection{Objective evaluations}

As objective evaluation indexes, we used root mean square error of log $F_0$: RMSE [Hz], unvoiced/voiced decision error: $U/V$ [\%], mel-cepstral distortion: MCD [dB], and log spectral distortion: LSD [dB]. We conducted all calculations after normalizing the power. For the calculation of RMSE when $F_0$ was transformed, we considered $F_0$ values extracted from natural speech multiplied by the scale factor as reference $F_0$ values.

\begin{table}[t]
\vspace{-7mm}
\caption{Results of objective evaluation.}
\vspace{-2mm}
\label{table:objective results}
\fontsize{8pt}{9.6pt}
\selectfont
{%
\begin{tabularx}{\columnwidth}{XYYYY}
\toprule
\multicolumn{1}{c}{model} & \multicolumn{1}{c}{RMSE} & \multicolumn{1}{c}{$U/V$} & \multicolumn{1}{c}{MCD} & \multicolumn{1}{c}{LSD} \\ \midrule
\multicolumn{1}{c}{} & \multicolumn{4}{c}{\cellcolor[HTML]{EFEFEF}$1.0 \times F_0$} \\
\multicolumn{1}{l}{PT-NSF} & 0.09 & 10 & 3.01 & 1.85 \\
\multicolumn{1}{l}{WORLD-NSF} & \textbf{0.06} & \textbf{9} & \textbf{2.70} & \textbf{1.69} \\
\multicolumn{1}{l}{QPPWG} & 0.08 & 11 & 3.07 & 1.83 \\
\multicolumn{1}{l}{uSFGAN (ours)} & 0.08 & 10 & 2.79 & 1.73 \\
\multicolumn{1}{l}{uSFGAN w/o $L_{reg}$} & 0.08 & 11 & 2.83 & 1.71 \\ \midrule
\multicolumn{1}{c}{} & \multicolumn{4}{c}{\cellcolor[HTML]{EFEFEF}$2.0 \times F_0$} \\
\multicolumn{1}{l}{PT-NSF} & 0.15 & 38 & 4.53 & -- \\
\multicolumn{1}{l}{WORLD-NSF} & 0.09 & \textbf{14} & 3.70 & -- \\
\multicolumn{1}{l}{QPPWG} & 0.33 & 36 & 4.26 & -- \\
\multicolumn{1}{l}{uSFGAN (ours)} & \textbf{0.06} & \textbf{14} & 3.81 & -- \\
\multicolumn{1}{l}{uSFGAN w/o $L_{reg}$} & 0.16 & 22 & \textbf{3.67} & -- \\ \midrule
\multicolumn{1}{c}{} & \multicolumn{4}{c}{\cellcolor[HTML]{EFEFEF}$0.5 \times F_0$} \\
\multicolumn{1}{l}{PT-NSF} & 0.69 & 24 & 3.57 & -- \\
\multicolumn{1}{l}{WORLD-NSF} & 0.68 & 55 & 3.60 & --\\
\multicolumn{1}{l}{QPPWG} & 0.17 & 37 & 3.45 & -- \\
\multicolumn{1}{l}{uSFGAN (ours)} & \textbf{0.14} & 40 & 3.08 & -- \\
\multicolumn{1}{l}{uSFGAN w/o $L_{reg}$} & 0.38 & \textbf{34} & \textbf{3.06} & -- \\
\bottomrule
\end{tabularx}%
}
\vspace{-3mm}
\end{table}

%

\begin{figure}[t]
\vspace{-7mm}
\begin{center}
    \includegraphics[width=7.9cm]{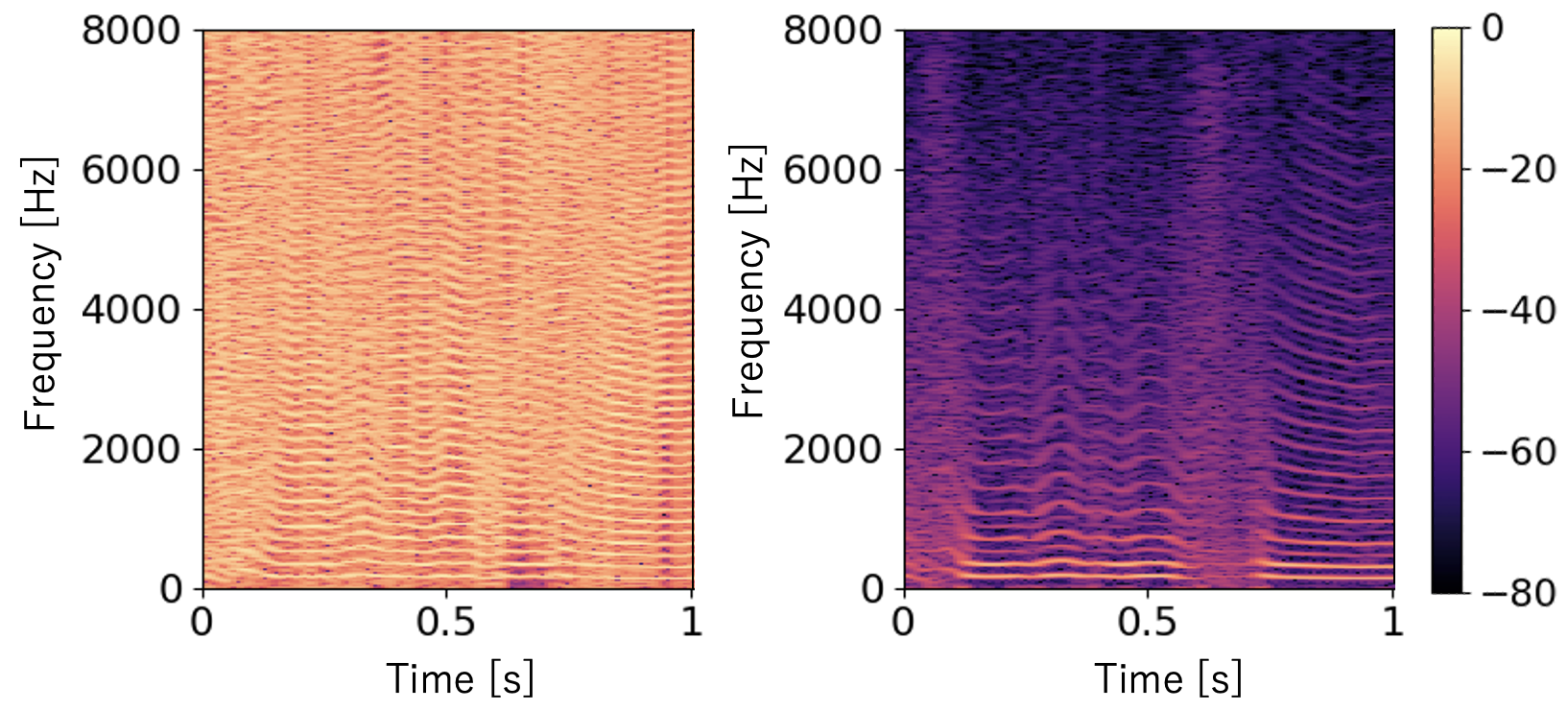}
    \vspace{-4mm}
    \caption{Spectrograms of source signals output from the source-networks of uSFGAN w/ (left) and w/o (right) $\mathcal{L}_{reg}$.}
    \label{fig:spectrogram}
    \vspace{1mm}
    
    \includegraphics[width=6.7cm]{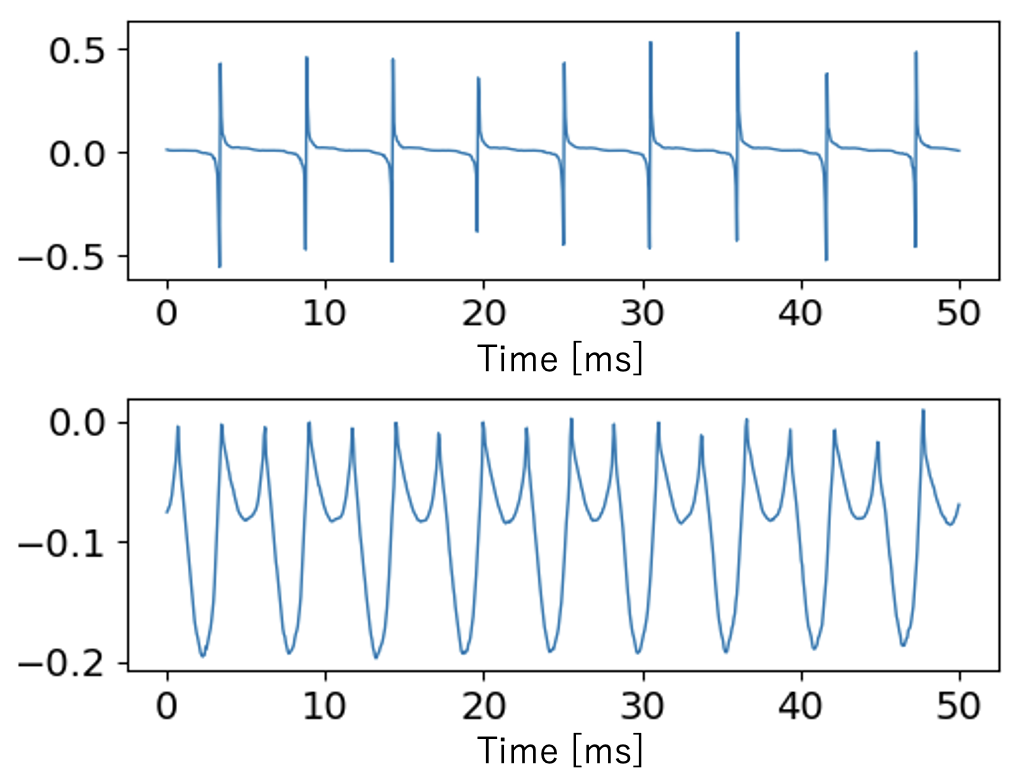}
    \vspace{-2mm}
    \caption{Source waveform output from the source-networks of uSFGAN w/ (top) and w/o (bottom) $\mathcal{L}_{reg}$. Given $F_0$ values over this segment are around 200 Hz.}
    \label{fig:waveform}
\end{center}
\vspace{-6mm}
\end{figure}

The objective evaluation results are shown in Table \ref{table:objective results}. It shows that uSFGAN tends to generate speech well conveying the information of the given auxiliary features with higher accuracy than PT-NSF, WORLD-NSF, and QPPWG. In addition, the spectral envelope regularization loss significantly improves the $F_0$ transformation accuracy in uSFGAN. An example of spectrograms and waveforms of the output source excitation signals from uSFGAN with and without the spectral envelope regularization loss are shown in Figs. \ref{fig:spectrogram} and \ref{fig:waveform}. We can find that the use of the spectral envelope regularization loss is effective for generating a reasonable source excitation signal of which spectral envelopes are flat and both harmonic components and periodic waveform shape well correspond to $F_0$ values.

\subsection{Subjective evaluations}

We conducted an opinion test on speech quality. Natural speech and synthetic speech from WORLD and three models: PT-NSF, QPPWG, and uSFGAN were evaluated by 10 subjects. 
We evaluated 160 utterances per each method per $F_0$ scaling factor. 
The synthetic speech was generated by scaling $F_0$ values by 1.0, 2.0 and 0.5 times.

As the experimental result, mean opinion scores (MOS) are shown in Table \ref{table:MOS}. It shows that uSFGAN significantly outperforms WORLD, PT-NSF, and QPPWG in both $1.0 \times F_0$ and $0.5 \times F_0$. In $2.0 \times F_0$, although WORLD still achieves the best speech quality, uSFGAN achieves comparable speech quality to PT-NSF and significantly better speech quality than QPPWG.

\begin{table}[t]
\caption{Speech quality MOS evaluations with 95\% CI.}
\vspace{-2mm}
\label{table:MOS}
\fontsize{8.0pt}{9.6pt}
\selectfont
{%
\begin{tabularx}{\columnwidth}{Xrrr}
\toprule
& \multicolumn{1}{c}{$1.0 \times F_0$} & \multicolumn{1}{c}{$2.0 \times F_0$} & \multicolumn{1}{c}{$0.5 \times F_0$} \\ \midrule
Natural & $4.58 \pm 0.18$ & \multicolumn{1}{c}{--}  & \multicolumn{1}{c}{--} \\
WORLD   & $3.93 \pm 0.25$ & $\textbf{2.71} \pm \textbf{0.25}$ & $2.66 \pm 0.27$ \\
PT-NSF     & $3.75 \pm 0.29$ & $2.21 \pm 0.27$ & $2.09 \pm 0.25$ \\
QPPWG   & $3.66 \pm 0.27$ & $1.48 \pm 0.21$ & $2.41 \pm 0.32$ \\
uSFGAN  & $\textbf{4.07} \pm \textbf{0.26}$ & $2.15 \pm 0.24$ & $\textbf{2.94} \pm \textbf{0.31}$ \\ \bottomrule
\end{tabularx}%
}
\vspace{-4mm}
\end{table}

To investigate the controllability, we conducted an ABX test to evaluate the perceptual accuracy of $F_0$ conversion by 10 subjects. 
We evaluated 100 utterances per each method per $F_0$ scaling factor. 
The synthetic speech by WORLD was used as a reference speech for three models: PT-NSF, QPPWG, and uSFGAN. The subjects selected a speech sample of which pitch was closer to that of the reference speech.

The experimental results are shown in Table \ref{table:ABX}. Each entry in the table shows the number of times selected as closer to the reference pitch in each pair comparison from the three models. It shows that uSFGAN significantly outperforms the other models in both $2.0 \times F_0$ and $0.5 \times F_0$. 
Our samples can be found on our demo page \cite{demopage}.

\begin{table}[tb]
\caption{Perceptive $F_0$ accuracy ABX evaluations with 95\% CI.}
\vspace{-2mm}
\label{table:ABX}
\fontsize{8.0pt}{9.6pt}
\selectfont
{%
\begin{tabularx}{\columnwidth}{Xrr}
\toprule
 & \multicolumn{1}{c}{$2.0 \times F_0$} & \multicolumn{1}{c}{$0.5 \times F_0$} \\ \midrule
PT-NSF / QPPWG & $\textbf{68} ~/~ 32 ~\pm 1.8 ~[\%]$ & $14 ~/~ \textbf{86} ~\pm 1.4 ~[\%]$ \\
QPPWG / uSFGAN & $7 ~/~ \textbf{93} ~\pm 1.0 ~[\%]$ & $16 ~/~ \textbf{84} ~\pm 1.4 ~[\%]$ \\
uSFGAN / PT-NSF & $\textbf{65} ~/~ 35 ~\pm 1.9 ~[\%]$ & $\textbf{95} ~/~ 5 ~\pm 0.9 ~[\%]$ \\ \bottomrule
\end{tabularx}%
}
\vspace{-5mm}
\end{table}

\section{Conclusions}

In this paper, we have proposed a unified neural vocoder framework based on the source-filter model, which is called unified source-filter GAN. In the proposed neural vocoder, a sinusoidal signal as the additional input, the pitch-dependent dilated convolution, and spectral envelope regularization loss have been implemented for factorizing the overall network into the source-network and the filter-network. The experimental evaluation results have demonstrated that the proposed neural vocoder can significantly improve $F_0$ transformation accuracy while achieving high speech quality. 
Thus, uSFGAN can be applied to entertaining TTS by combining it with an acoustic model that outputs vocoder features.

\section{Acknowledgements}

This work was supported in part by JST, CREST, and JPMJCR19A3.


\bibliographystyle{IEEEtran}

\bibliography{template}

\end{document}